\documentclass[sigconf]{acmart}

\usepackage{listings}
\usepackage{xcolor}

\definecolor{codegreen}{rgb}{0,0.6,0}
\definecolor{codegray}{rgb}{0.5,0.5,0.5}
\definecolor{codepurple}{rgb}{0.58,0,0.82}
\definecolor{backcolour}{rgb}{0.95,0.95,0.92}

\lstdefinestyle{mystyle}{
    backgroundcolor=\color{yellow!10},   
    commentstyle=\color{codegreen},
    keywordstyle=\color{magenta},
    numberstyle=\tiny\color{codegray},
    stringstyle=\color{codepurple},
    basicstyle=\ttfamily\footnotesize,
    breakatwhitespace=false,         
    breaklines=true,                 
    captionpos=b,                    
    keepspaces=true,                 
    numbers=left,                    
    numbersep=5pt,                  
    showspaces=false,                
    showstringspaces=false,
    showtabs=false,                  
    tabsize=2,
    xleftmargin=9pt,
    xrightmargin=3pt,
}
\PassOptionsToPackage{bookmarks={false}}{hyperref}
\hypersetup{draft}
\AtBeginDocument{%
  \providecommand\BibTeX{{%
    \normalfont B\kern-0.5em{\scshape i\kern-0.25em b}\kern-0.8em\TeX}}}

\setcopyright{acmcopyright}
\copyrightyear{2021} 
\acmYear{2021} 
\setcopyright{acmcopyright}\acmConference[ASPDAC '21]{26th Asia and South Pacific Design Automation Conference}{January 18--21, 2021}{Tokyo, Japan}
\acmBooktitle{26th Asia and South Pacific Design Automation Conference (ASPDAC '21), January 18--21, 2021, Tokyo, Japan}
\acmPrice{15.00}
\acmDOI{10.1145/3394885.3431629}
\acmISBN{978-1-4503-7999-1/21/01}

\begin{document}

\title{When Machine Learning Meets Quantum Computers:\\ A Case Study}
\subtitle{(Invited Paper)}
\author{Weiwen Jiang}
\email{wjiang2@nd.edu}
\affiliation{%
  \institution{University of Notre Dame}
  \streetaddress{P.O. Box 1212}
  \city{Notre Dame}
  \state{IN}
  \country{U.S.}
  \postcode{43017}
}

\author{Jinjun Xiong}
\email{jinjun@us.ibm.com}
\affiliation{%
  \institution{{\normalsize IBM Thomas J. Watson Research Center}}
  \streetaddress{1 Th{\o}rv{\"a}ld Circle}
  \city{Yorktown Heights}
  \state{NY}
  \country{U.S.}}

\author{Yiyu Shi}
\email{yshi4@nd.edu}
\affiliation{%
  \institution{University of Notre Dame}
  \streetaddress{}
   \city{Notre Dame}
  \state{IN}
  \country{U.S.}
  \postcode{43017}}

\renewcommand{\shortauthors}{W. Jiang, et al.}
\renewcommand{\shorttitle}{When Machine Learning Meets Quantum Computers: A Case Study}

\newcommand{\rtodo}[1]{\textcolor[rgb]{0.00,0.00,0.00}{#1}}
\newcommand{\todo}[1]{\textcolor[rgb]{0.00,0.00,0.00}{#1}}
\newcommand{\app}[1]{\textcolor[rgb]{0.00,0.00,1.00}{#1}}
\newcommand{\revision}[1]{\textcolor[rgb]{0.00,0.00,0.00}{#1}}


\begin{abstract}
Along with the development of AI democratization, the machine learning approach, in particular neural networks, has been applied to wide-range applications.
In different application scenarios, the neural network will be accelerated on the tailored computing platform. 
The acceleration of neural networks on classical computing platforms, such as CPU, GPU, FPGA, ASIC, has been widely studied; however, when the scale of the application consistently grows up, the memory bottleneck becomes obvious, widely known as memory-wall. 
In response to such a challenge, advanced quantum computing, which can represent $2^N$ states with $N$ quantum bits (qubits), is regarded as a promising solution.
It is imminent to know how to design the quantum circuit for accelerating neural networks.
Most recently, there are initial works studying how to map neural networks to actual quantum processors. 
To better understand the state-of-the-art design and inspire new design methodology, this paper carries out a case study to demonstrate an end-to-end implementation.
On the neural network side, we employ the multilayer perceptron to complete image classification tasks using the standard and widely used MNIST dataset.
On the quantum computing side, we target IBM Quantum processors, which can be programmed and simulated by using IBM Qiskit.
This work targets the acceleration of the inference phase of a trained neural network on the quantum processor.
Along with the case study, we will demonstrate the typical procedure for mapping neural networks to quantum circuits. 
\end{abstract}

\begin{CCSXML}
<ccs2012>
 <concept>
  <concept_id>10010520.10010553.10010562</concept_id>
  <concept_desc>Computer systems organization~Embedded systems</concept_desc>
  <concept_significance>500</concept_significance>
 </concept>
 <concept>
  <concept_id>10010520.10010575.10010755</concept_id>
  <concept_desc>Computer systems organization~Redundancy</concept_desc>
  <concept_significance>300</concept_significance>
 </concept>
 <concept>
  <concept_id>10010520.10010553.10010554</concept_id>
  <concept_desc>Computer systems organization~Robotics</concept_desc>
  <concept_significance>100</concept_significance>
 </concept>
 <concept>
  <concept_id>10003033.10003083.10003095</concept_id>
  <concept_desc>Networks~Network reliability</concept_desc>
  <concept_significance>100</concept_significance>
 </concept>
</ccs2012>
\end{CCSXML}


\keywords{neural networks, MNIST dataset, quantum computing, IBM Quantum, IBM Qiskit}

%   \includegrapsics[width=\textwidth]{sampleteaser}

\maketitle

\section{Introduction}
In the past few years, we have witnessed many breakthroughs in both machine learning and quantum computing research fields.
On machine learning, the automated machine learning (AutoML) \cite{zoph2016neural,zoph2018learning} significantly reduces the cost of designing neural networks to achieve AI democratization.
On quantum computing, the scale of the actual quantum computers has been rapidly evolving (e.g., IBM \cite{ibm2020roadmap} recently announced to debut quantum \todo{computer} with 1,121 quantum bits (\revision{qubits}) in 2023).
Such two research fields, however, have met the bottlenecks when applying the theoretical knowledge in practice.
With the large-size inputs, the size of machine learning models (i.e., neural networks) significantly exceed the resource provided by the classical computing platform (e.g., GPU and FPGA); on the other hand, the development of quantum applications is far behind the development of quantum hardware, that is, it lacks killer applications to take full advantage of high-parallelism provided by a quantum computer.
As a result, it is natural to see the emerging of a new research field, quantum machine learning. 

Like applying machine learning to the classical hardware accelerators, when machine learning meets quantum computers, there will be tons of opportunities along with the challenges.
The development of machine leering on the classical hardware accelerator experienced two phases: (1) the design of neural network tailored hardware \cite{zhang2015optimizing,jiang2018heterogeneous,zhang2018dnnbuilder,jiang2019achieving,li2016using,li2017neural}, and (2) the co-design of neural network and hardware accelerator \cite{jiang2019accuracy,jiang2019integrating,yang2020coasp,bian2020nass,jiang2020device,ding2020hardware,wu2019fbnet,cai2018proxylessnas,tan2019mnasnet,hao2019fpga,hao2019nais,zeng2020towards,wu2020enabling}.
To best exploit the power of the quantum computer, it would be essential to conduct the co-design of neural network and quantum circuits design; however, with the different basic logic gates between quantum circuit and classical circuit designs, it is still unclear how to design a quantum accelerator for the neural network.

In this work, we aim to fix such a missing link by providing an open-source design framework.
In general, the full acceleration system will be divided into three parts, the data pre-processing and data post-processing on a classical computer, and the neural network accelerator on the quantum circuit. 
In the quantum circuit, it will further include the quantum state preparation and the quantum computing-based neural computation.
In the following of this paper, we will introduce all the above components in detail and demonstrate the implementation using IBM Qiskit for quantum circuit design and Pytorch for the machine learning model process. 

The remainder of the paper is organized as follows. Section 2 presents an overview of the full system. Section 3 presents the case study on the MNIST dataset.
% accurate analytic model and novel XFER design in Super-LIP.
Insights are discussed in Section 4.
% Section 6 discusses related work.
Finally, concluding remarks are given in Section 5.

\section{Overview}

\begin{figure}[t]
\centering
\includegraphics[width=0.99\linewidth]{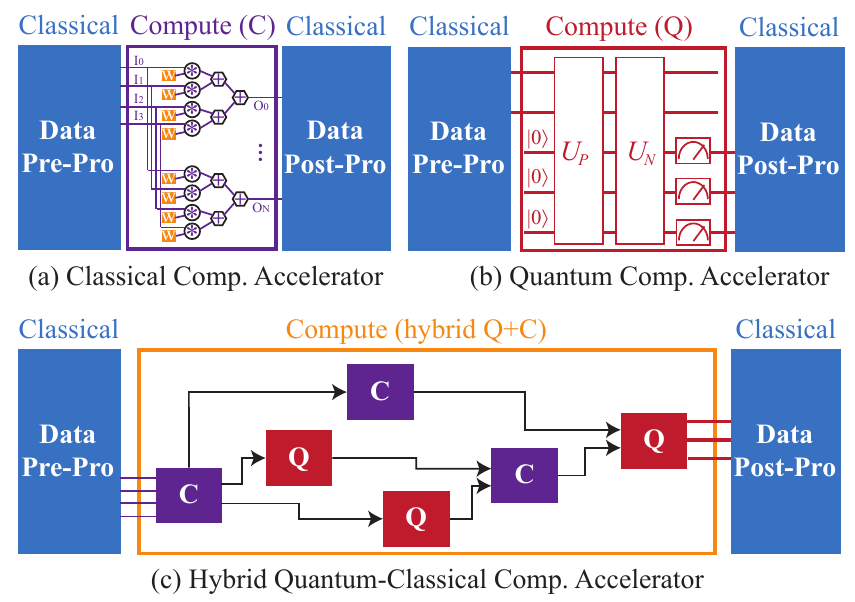}
\caption{Illustration of three different types of computing schemes: (a) classical computing ``C'' based neural computation, where $W$ is weights; (b) quantum computing ``Q'' based neural computation, where $U_p$ is the quantum-state preparation and $U_N$ is the neural computation; (c) hybrid quantum-classical computing ``Q+C'' based neural computation.}
\label{fig:schemes}
\end{figure}

Figure \ref{fig:schemes} demonstrates three types of neural network design: (1) the classical hardware accelerator; (2) the pure quantum computing based accelerator; (3) the hybrid quantum and classical accelerator.
All of these accelerators follow the same flow that the data will be first pre-processed, then the neural computation is accelerated, and finally, the output data will go through the post-processing to obtain the final results.
 
% Although the research on neural networks in quantum computing can trace back to the 1990s  \citep{kak1995quantum,purushothaman1997quantum,ezhov2000quantum},
% but only recently, along with the revolution of quantum computers, the implementation of neural networks on actual quantum \todo{computer} emerges \citep{francesco2019artificial,jiang2020co,bisarya2020breast}. There are mainly three different directions to exploit the power of quantum computers:
% (1) applying the Quantum Random Access Memory (QRAM) \citep{blencowe2010quantum}; (2) employing pure quantum \todo{computers}; (3) bridging different platforms for a hybrid quantum-classical computing \citep{mcclean2016theory}.

% \subsection{QRAM-based implementation}

% \cite{kerenidis2019quantum} is a typical work to implement neural networks with QRAM.
% Using QRAM provides the highest flexibility, such as implementing non-linear functions using lookup tables.
% But QRAM itself has limitations: instead of using the widely applied superconducting \revision{qubits} \citep{arute2019quantum,ibm2016ibmq}, QRAM needs the support of spin qubit \citep{veldhorst2015two} to provide relatively long lifetime.
% To make the system practical, there is still a long way to go.

\subsection{Classical acceleration}
After the success of deep neural networks (e.g., Alexnet \cite{krizhevsky2017imagenet} and VGGNet \cite{simonyan2014very}) in achieving high accuracy, designing hardware accelerator became the hot topic in accelerating the execution of deep neural networks.
On the application-specific integrated circuit (ASIC), works \cite{du2015shidiannao,zhang2018thundervolt,zhang2018fate,zhang2019compact,chen2016eyeriss} studied how to design neural network accelerator using different dataflows, including weight stationery, output stationery, etc. By selecting dataflow for a dedicated neural computation, it can maximize the data reuse to reduce the data movement and accelerate the process, which derived the co-design of neural network and ASICs \cite{yang2020co}.

On the FPGA, work \cite{zhang2015optimizing} first proposed the tiling based design to accelerate the neural computation, and works \cite{jiang2018heterogeneous,zhang2018dnnbuilder,jiang2019achieving,li2015fpga} gave different designs and extended the implementation to multiple FPGAs.
Driven by the AutoML, work \cite{jiang2019accuracy} proposed the first co-design framework to involve the FPGA implementation into the search loop, so that both software accuracy and hardware efficiency can be maximized.
The co-design philosophy also applied in other designs \cite{zhang2019neural,jiang2020hardware,hao2019fpga,hao2019nais} and in this direction, there exist many research works in further integrating the model compression into consideration \cite{lu2019neural,jiang2020standing}, accelerating the search process \cite{li2020edd,zhang2020dna},

\subsection{Pure quantum computing}

Most recently, the emerging works in using the quantum circuit to accelerate neural computation.
The typical work include \citep{francesco2019artificial,tacchino2020quantum,jiang2020co}, among which the work \cite{jiang2020co} first demonstrates the potential quantum advantage that can be achieved by using a co-design philosophy.
These works encode data to either \revision{qubits} \citep{francesco2019artificial} or \revision{qubit} states \citep{jiang2020co} and use superconducting-based quantum \todo{computers} to run neural networks.
These methods have the following limitations:
Due to the short decoherence times in the superconducting-based quantum \todo{computers}, the condition logic is not supported in the computing process.
This makes it hard to implement a function that is not differentiable at all points, like the commonly used Rectified Linear Unit (ReLU) in machine learning models.
% non-linear functions such as the most 
However, it also has advantages, such as the design can be directly evaluated on an actual quantum computer, and there is no communication between the quantum-classical interface during the computation.

In the quantum circuit design, it includes two components: $U_P$ for quantum states preparation and $U_N$ for neural computation, as shown in Figure \ref{fig:schemes}(b).
After the component $U_N$, it will measure the quantum qubits to extract the output data, which will be further sent to the data post-processing unit to obtain the final results.

\subsection{Hybrid quantum-classical computing}

To overcome the disadvantage of pure quantum computing and take full use of classical computing, the hybrid quantum-classical computing for machine learning tasks is proposed \cite{broughton2020tensorflow}.
It establishes a computing paradigm where different neurons can be implemented on either quantum or classical \todo{computers}, as demonstrated in Figure \ref{fig:schemes}(c).
\revision{This brings flexibility in implementing functions (e.g., ReLU). However, at the same time, it will lead to massive data transfer between quantum and classical {computers}.}

\subsection{Our Focus in The Case Study}
This work focus on providing a full workflow, starting from the data pre-processing, going through quantum computing acceleration, and ending with the data post-processing. We will apply the MNIST data set as an example to carry out a case study.

Computing architecture and neural operation can affect the design.
In this work, for the computing architecture, we focus on the pure quantum computing design, since it can be easily extended to the hybrid quantum-classical design by connecting the inputs and output of the quantum acceleration to the traditional classical accelerator; for the neural network, we focus on the multi-layer perceptron, which is the basic operation for a large number of neural computation, like the convolution.

\section{Case Study on MNIST Dataset}

In this section, we will demonstrate the detailed implementation of four components in the pure quantum computing based neural computation as shown in Figure \ref{fig:schemes}(b): data pre-processing, quantum state preparation ($U_P$), neural computation ($U_N$), and data post-processing. 

\subsection{Data Pre-Processing}

The first step of the whole procedure is to prepare the quantum data to be encoded to the quantum states.
Kindly note in order to utilize $N$ qubits to represent $2^N$ data, it has constraints on the numbers; more specifically, if a vector $U_0$ of $2^N$ data can be arranged in the first column of a unitary matrix $U$, then for the initial state of $|\psi\rangle=1\cdot |0\rangle^{\otimes N}$, we can obtain $U_0$ by conducting $U|\psi\rangle=U_0$, where $|0\rangle^{\otimes N}$ represents the zero state with $N$ qubits.

\begin{lstlisting}[label=code:preproc,language=Python, 
caption=Converting classical data to qautnum data,frame=single
]
import torch
import numpy as np
import torchvision.transforms as transforms
# Input: img_size=4 to represent the resolution of 4*4 
class ToQuantumData(object):
    def __call__(self, tensor):
        device = torch.device("cuda" if torch.cuda.is_available() else "cpu")
        data = tensor.to(device)
        input_vec = data.view(-1)
        vec_len = input_vec.size()[0]
        input_matrix = torch.zeros(vec_len, vec_len)
        input_matrix[0] = input_vec
        input_matrix = np.float64(input_matrix.transpose(0,1))
        u, s, v = np.linalg.svd(input_matrix)
        output_matrix = torch.tensor(np.dot(u, v))
        output_data = output_matrix[:, 0].view(1, img_size,img_size)
        return output_data
# Similarly, we have "class ToQuantumMatrix(object)" which return the output_matrix
transform = transforms.Compose([transforms.Resize((img_size, img_size)), transforms.ToTensor(), ToQuantumData()])
# transform = transforms.Compose([transforms.Resize((img_size,img_size)),transforms.ToTensor(),transforms.Normalize((0.1307,), (0.3081,)), ToQuantumData()])
\end{lstlisting}

Listing \ref{code:preproc} demonstrates the data conversion from the classical data to quantum data.
We utilize the transforms in torchvision to complete the data conversation.
More specifically, we create the ToQuantumData class in Line 5.
It will receive a tensor (the original data) as input (Line 6).
We apply Singular Value Decomposition (svd) provided by np.linalg to obtain the unitary matrix output\_matrix (Line 14), then we extract the first vector from output\_matrix as the output\_data  (Line 16), where the output\_matrix represents $U$ and the output\_data represents $U_0$.
After we build the ToQuantumData class, we will integrate it into one ``transform'' variable, which can further include the data pre-processing functions, such as image resize (Line 20) and data normalization (Line 21).
In creating the data loader, we can apply the ``transform'' to the dataset (e.g., we can obtain train data by using ``train\_data=datasets.MNIST(root=datapath, train=True,download=True, transform=transform)'').

% Specifically, in order to encode the classical data to quantum data, we

% \begin{lstlisting}[language=Python, caption=Python example,frame=single]
% import numpy as np
    
% def incmatrix(genl1,genl2):
%     m = len(genl1)
%     n = len(genl2)
%     M = None #to become the incidence matrix
%     VT = np.zeros((n*m,1), int)  #dummy variable
    
%     #compute the bitwise xor matrix
%     M1 = bitxormatrix(genl1)
%     M2 = np.triu(bitxormatrix(genl2),1) 

%     for i in range(m-1):
%         for j in range(i+1, m):
%             [r,c] = np.where(M2 == M1[i,j])
%             for k in range(len(r)):
%                 VT[(i)*n + r[k]] = 1;
%                 VT[(i)*n + c[k]] = 1;
%                 VT[(j)*n + r[k]] = 1;
%                 VT[(j)*n + c[k]] = 1;
                
%                 if M is None:
%                     M = np.copy(VT)
%                 else:
%                     M = np.concatenate((M, VT), 1)
                
%                 VT = np.zeros((n*m,1), int)
    
%     return M
% \end{lstlisting}

\subsection{$U_P$: Quantum State Preparation}
Theoretically, with the $n\times n$ unitary matrix $U$, we can directly operate the oracle on the quantum circuit to change $2^N$ states from the zero state $|0\rangle^{\otimes N}$ to $U_0$. 
This process is widely known as quantum-state preparation.
The efficiency of quantum-state preparation can significantly affect the complexity of the whole circuit, and therefore, it is quite important to improve the efficiency of such a process.
In general, there are two typical ways to perform the quantum-state preparation: (1) quantum random access memory (qRAM) \cite{lvovsky2009optical} based approach \cite{allcock2020quantum,kerenidis2016quantum} and (2) computing based approach \cite{sanders2019black,grover2000synthesis,bausch2020fast}.
Let's first see the qRAM-based approach, where the vector in $U_0$ will be stored in a binary-tree based structure in qRAM, which can be queried in quantum superposition and can generate the states efficiently.
In IBM Qiskit, it provides the initialization function to perform quantum-state preparation, which is based on the method in \cite{shende2006synthesis}.

\begin{lstlisting}[label=code:qs-prep,language=Python, 
caption=Quantum-State Preparation in IBM Qiskit,frame=single
]
from qiskit import QuantumRegister, QuantumCircuit, ClassicalRegister
from qiskit.extensions import XGate, UnitaryGate
from qiskit import Aer, execute
import qiskit
# Input: a 4*4 matrix (data) holding 16 input data
inp = QuantumRegister(4,"in_qbit")
circ = QuantumCircuit(inp)
data_matrix = Q_InputMatrix = ToQuantumMatrix()(data.flatten())
circ.append(UnitaryGate(data_matrix, label="Input"), inp[0:4])
# Using StatevectorSimulator from the Aer provider
simulator = Aer.get_backend('statevector_simulator')
result = execute(circ, simulator).result()
statevector = result.get_statevector(circ)
print(statevector)
\end{lstlisting}

In Listing \ref{code:qs-prep}, we give the codes to initialize the quantum states, using the unitary matrix $U$ which is converted from the original data in Listing \ref{code:preproc}(see Line 18).
In this code snippet, we first create a 4-qubit QuantumRegister ``inp'' (line 6) and the quantum circuit (line 7).
Then, we convert the input data to data\_matrix, which is then employed to initialize the circuit using function UnitaryGate from qiskit.extensions.
Finally, from line 10 to line 14, we output the states of all qubits to verify the correctness.

\subsection{$U_N$: Neural Computation}
Now, we have encoded the image data (16 inputs) onto 4 qubits.
The next step is to perform the neural computation, that is, the weighted sum with quadratic function using the given binary weights $W$.
Neural computation is the key component in quantum machine learning implementation.
To clearly introduce this component, we first consider the computation of the hidden layer, which can be further divided into two stages: (1) multiplying inputs and weights, and (2) applying the quadratic function on the weighted sum.
Then, we will present the computation of the output layer to obtain the final results.

\vspace{5pt}
\noindent\textbf{Computation of one neural in the hidden layer}

\textit{Stage 1: multiplying inputs and weights.}
Since the weight $W$ is given, it is pre-determined.
We use the quantum gate to operate the weights with the inputs. The quantum gates applied here include the $X$ gate and the 3-controlled-Z gate with 3 trigger qubits.
The function of such a 3-controlled-Z is to flip the sign of state $|1111\rangle$, and the function of $X$ gate is to swap one state to another state.

For example, if the weight for state $|0011\rangle$ is $-1$.
We operate it on the input follows three steps.
First, we swap the amplitude of state $|0011\rangle$ to state $|1111\rangle$ using two $X$ gates on the first two qubits.
Then, in the second step, we apply controlled-Z gate to flip the sign of the state $|1111\rangle$.
Finally, in the third step, we swap the amplitude of state $|1111\rangle$ back to state $|0011\rangle$ using two $X$ gates on the first two qubits.
Therefore, we can transverse all weights and apply the above three steps to flip the sign of corresponding states.
Kindly note that since the non-linear function is a quadratic function, if the number of $-1$ is larger than $+1$, we can flip all signs of weights to minimize the number of gates to be put in the circuit.

\begin{lstlisting}[label=code:qs-mulwei,language=Python, 
caption=Multiplying inputs and weights on quantum,frame=single
]
def cccz(circ, q1, q2, q3, q4, aux1, aux2):
    # Apply Z-gate to a state controlled by 4 qubits
    circ.ccx(q1, q2, aux1)
    circ.ccx(q3, aux1, aux2)
    circ.cz(aux2, q4)
    # cleaning the aux bits
    circ.ccx(q3, aux1, aux2)
    circ.ccx(q1, q2, aux1)
    return circ
def neg_weight_gate(circ,qubits,aux,state):
    for idx in range(len(state)):
        if state[idx]=='0':
            circ.x(qubits[idx])
    cccz(circ,qubits[0],qubits[1],qubits[2],qubits[3],aux[0],aux[1])
    for idx in range(len(state)):
        if state[idx]=='0':
            circ.x(qubits[idx])
# input: weight vector, weight_1_1
aux = QuantumRegister(2,"aux_qbit")
circ.add_register(aux)
if weight_1_1.sum()<0:
    weight_1_1 = weight_1_1*-1
for idx in range(weight_1_1.flatten().size()[0]):
    if weight_1_1[idx]==-1:
        state = "{0:b}".format(idx).zfill(4)
        neg_weight_gate(circ,inp,aux,state)
        circ.barrier()
print(circ)
\end{lstlisting}

Listing \ref{code:qs-mulwei} demonstrates the procedure of multiplying inputs and weights.
In the list, the function cccz utilizing the basic quantum logic gates to realize the 3-controlled-Z gate with 3 control qubits.
The involved basic gates include Toffoli gate (i.e., CCX) and controlled-Z gate (i.e., CZ).
Since such a function needs auxiliary (a.k.a., ancilla) qubits, we include 2 additional qubits (i.e., $aux$) in the quantum circuit (i.e., $circ$), as shown in Lines 19-20. 

The function neg\_weights\_gate flips the sign of the given state, applying the 3-step process. Lines 11-13 complete the first step to swap the amplitude of the given state to the state of $|1\rangle^{\otimes 4}$.
Then, the cccz gate is applied to complete the second step.
Finally, from line 15 to line 17, the amplitude is swap back to the given state.

With the above two functions, we traverse the weights to assign the sign to each state from Lines 21-27. 
Kindly note that, after this operation, the states vector changed from the initial state $|\psi\rangle=U_0$ to $|\psi^{\prime}\rangle=U_0^{\prime}$ where the states have the weights.

\textit{Stage 2: applying a quadratic function on the weighted sum.}
In this stage, it also follows 3 steps to complete the function.
In the first step, we apply the Hadamard (H) gates on all qubits to accumulates all states to the zero states. 
Then, the second step swap the amplitude of zero state $|0\rangle^{\otimes N}$ and the one-state $|1\rangle^{\otimes N}$.
Finally, the last step applies the N-control-X gate to extract the amplitude to one output qubit $O$, in which the probability of $O=|1\rangle$ is equal to the square of the weighted sum.

In the first step, the H gates can be applied to accumulate the amplitude of states, because the first row of $H^{\otimes 4}$ is $\frac{1}{4}\times[1,1,1,1]$ and the $H^{\otimes 4}|\psi^{\prime}\rangle$ performs the multiplication between the $4\times 4$ matrix and the state vector $U_0^{\prime}$.
As a result, the amplitude of $|0\rangle^{\otimes N}$ will be the weighted sum with the coefficient of $\frac{1}{\sqrt{N}}$.

\begin{lstlisting}[label=code:qs-quadr,language=Python, 
caption=Applying quadratic function on the weighted sum,frame=single
]
def ccccx(circ, q1, q2, q3, q4, q5, aux1, aux2):
    circ.ccx(q1, q2, aux1)
    circ.ccx(q3, q4, aux2)
    circ.ccx(aux2, aux1, q5)
    # cleaning the aux bits
    circ.ccx(q3, q4, aux2)
    circ.ccx(q1, q2, aux1)
    return circ
# input: circ after stage 1
hidden_neuron = QuantumRegister(1,"out_qbit")
circ.add_register(hidden_neuron)
circ.h(inp)
circ.x(inp)
ccccx(circ,inp[0],inp[1],inp[2],inp[3],hidden_neuron,aux[0],aux[1])
\end{lstlisting}

Listing \ref{code:qs-quadr} demonstrates the implementation of the quadratic function on the weighted sum on Qiskit.
In the list, function ccccx is based on the basic Toffoli gate (i.e., CCX) to implement a 4-control-X gate to swap the amplitude between the zero state $|0\rangle^{\otimes 4}$ and the one-state $|1\rangle^{\otimes 4}$.
In Line 14, $hidden\_neuron$ is an additional output qubit in the quantum circuit (i.e., $circ$) to hold the result for the neural computation, which is added in Lines 10-11. 

For a neural network with $N$ neurons in the hidden layer, it has $N$ sets of weights. 
We can apply the above neural computation on $N$ set of weights to obtain $N$ output qubits. 

\vspace{5pt}
\noindent\textbf{Computation of one neuron in the output layer}

With these $N$ output qubits, we have two choices: (1) go to the classical computer and then encode the output of these $N$ outputs to $\log_2{N}$ qubits and then repeat these computations for the hidden layer to obtain the final results; (2) continuously use these qubits to directly compute the outputs, but the fundamental computation needs to be changed to the multiplication between random variables because the data associated with a qubit represents the probability of the qubit to be $|0\rangle$ state.

In the following, we demonstrate the implementation of the second choices (fundamental details please refer to \cite{jiang2020co,tacchino2020quantum}).
In this example, we follow the network structure with 2 neurons in the hidden layer.
In addition, we consider there is only one parameter for the normalization function using one additional qubit for each output neuron.
Let $hidden\_neurons$ be the outputs of 2 neurons in the hidden layer; let $weight\_2\_1$ be the weights for the $1^{st}$ output neuron in the $2^{nd}$ layer; let norm\_flag\_1 and norm\_para\_1 be the normalization related parameters for the $1^{st}$ output neuron.
Then, we have the following implementation.

\begin{lstlisting}[label=code:sec-lyr,language=Python, 
caption=Implementation of the second layer neural computation without measurement after the first layer,frame=single
]
# Additional registers
inter_q_1 = QuantumRegister(1,"inter_q_1_qbits")
norm_q_1 = QuantumRegister(1,"norm_q_1_qbits")
out_q_1 = QuantumRegister(1,"out_q_1_qbits")
circ.add_register(inter_q_1,norm_q_1,out_q_1)
circ.barrier()
# Input and weight multiplication
if weight_2_1.sum()<0:
    weight_2_1 = weight_2_1*-1
idx = 0
for idx in range(weight_2_1.flatten().size()[0]):
    if weight_2_1[idx]==-1:
        circ.x(hidden_neurons[idx])
circ.h(inter_q_1)
circ.cz(hidden_neurons[0],inter_q_1)
circ.x(inter_q_1)
circ.cz(hidden_neurons[1],inter_q_1)
circ.x(inter_q_1)
# quadratic function on weighted sum
circ.h(inter_q_1)
circ.x(inter_q_1)
circ.barrier()
# normalization for two cases
norm_init_rad = float(norm_para_1.sqrt().arcsin()*2)
circ.ry(norm_init_rad,norm_q_1)
if norm_flag_1:
    circ.cx(inter_q_1,out_q_1)
    circ.x(inter_q_1)
    circ.ccx(inter_q_1,norm_q_1,out_q_1)
else:
    circ.ccx(inter_q_1,norm_q_1,out_q_1)
# Recove the inputs for the next neuron computation
for idx in range(weight_2_1.flatten().size()[0]):
    if weight_2_1[idx]==-1:
        circ.x(hidden_neurons[idx])
\end{lstlisting}

In the above list, it follows the 2-stage pattern for the computation in the hidden layer.
If we modify all sub-index $\_1$ to $\_2$, then we can obtain the quantum circuit for the second output neuron.

\subsection{Data Post-Processing}
After all outputs are computed and stored in the out\_q\_1 and out\_q\_2 qubits, we can then measure the output qubits, run a simulation or execute on the IBM Q processors, and finally obtain the classification as follows.

\begin{lstlisting}[label=code:res,language=Python, 
caption=Extract the classification results,frame=single
]
from qiskit.tools.monitor import job_monitor

def fire_ibmq(circuit,shots,Simulation = False,backend_name='ibmq_essex'):     
    count_set = []
    if not Simulation:
        provider = IBMQ.get_provider('ibm-q-academic')
        backend = provider.get_backend(backend_name)
    else:
        backend = Aer.get_backend('qasm_simulator')
    job_ibm_q = execute(circuit, backend, shots=shots)
    job_monitor(job_ibm_q)
    result_ibm_q = job_ibm_q.result()
    counts = result_ibm_q.get_counts()
    return counts

def analyze(counts):
    mycount = {}
    for i in range(2):
        mycount[i] = 0
    for k,v in counts.items():
        bits = len(k) 
        for i in range(bits):            
            if k[bits-1-i] == "1":
                if i in mycount.keys():
                    mycount[i] += v
                else:
                    mycount[i] = v
    return mycount,bits

qc_shots=8192
counts = fire_ibmq(circ,qc_shots,True)
(mycount,bits) = analyze(counts)
class_prob=[]
for b in range(bits):
    class_prob.append(float(mycount[b])/qc_shots)
class_prob.index(max(class_prob))
\end{lstlisting}

Listing \ref{code:res} demonstrate the above three tasks. 
The fire\_ibmq function can execute the constructed circuit in either simulation or a given IBM Q processor backend.
The parameter ``shots'' defines the number of execution to be executed.
Finally, the counts for each state will be returned.
On the implementation, the probability of each qubit (instead of each state) gives the probability to choose the corresponding class.
Therefore, we create the ``analyze'' function to get the probability for each qubits.
Finally, we obtain the classification results by extracting the index of the max probability in the ``class\_prob'' set.

Kindly note that the Listing \ref{code:res} can also be applied for the hybrid quantum-classical computing.

% \clearpage
% \section{Optimizing Neural Computation on Quantum Circuits}

\section{Insights}

From the study of implementing neural networks onto the quantum circuits, there are several insights in terms of achieving quantum advantages, listed as follows.

\begin{itemize}
\item \textbf{Data encoding:} this case study encodes $2^N$ data to $N$ quantum qubits, which provides the opportunity to achieve quantum advantage for conducting inference for each input. An alternative way is to encode $N$ data to $N$ qubits, however, with the consideration that each data needs to be operated in the neural computation, such an encoding approach can hardly achieve the quantum advantage. 
\item \textbf{Quantum-state preparation:} by encoding $2^N$ data to $N$ quantum qubits, we can achieve quantum advantage only if the quantum-state preparation can be efficiently conducted with complexity at $O(N)$. 
\item \textbf{Quantum computing-based neural computation:} Neural computation can also become the performance bottleneck, using the design in Listing \ref{code:qs-mulwei} to flip one sign at each time, it requires $O(2^N)$ gates in the worst case. To overcome this, \cite{jiang2020co} proposed a co-design approach to reduce the number of gates to $O(N^2)$.
\end{itemize}

\section{Conclusion}

This work demonstrates the framework in implementing neural networks onto quantum circuits.
It is composed of three main components, including data pre-processing, neural computation acceleration, and data post-processing. Based on such a working flow, the data will be first encoded to quantum states and then operated to complete the operations in a neural network.
The source codes can be found in \url{https://github.com/weiwenjiang/QML_tutorial}

\section*{Acknowledgements}
This work is partially supported by IBM
and University of Notre Dame (IBM-ND) Quantum program, and in part by the IBM-ILLINOIS Center for Cognitive Computing Systems Research.

% % \clearpage
% \bibliographystyle{ACM-Reference-Format}
% \bibliography{ref}


\end{document}